\RequirePackage{fix-cm}
\documentclass[twocolumn,epjc3,nofootinbib]{svjour3}   
\usepackage[T1]{fontenc} 
\usepackage{graphicx}
\usepackage{subfigure}
\usepackage{amsmath} 

\usepackage[colorlinks=true,citecolor=blue,urlcolor=magenta,breaklinks]{hyperref}
\usepackage{natbib}
\usepackage{footnote}

\journalname{Eur. Phys. J. C}

\begin{document}

\title{The spectroscopy of solar sterile neutrinos}
\author{Il\'idio Lopes~\thanksref{addr1,addr2,e1}}
\thankstext{e1}{e-mail: ilidio.lopes@tecnico.ulisboa.pt}
\institute{Centro de Astrof\'{\i}sica e Gravita\c c\~ao  - CENTRA,
Departamento de F\'{\i}sica, Instituto Superior T\'ecnico - IST,\\
Universidade de Lisboa - UL, Av. Rovisco Pais 1, 1049-001 Lisboa, Portugal\label{addr1}
\and
Institut d'Astrophysique de Paris (UMR 7095: CNRS \& UPMC, Sorbonne Universit\'es), \\
98 bis Bd Arago, F-75014 Paris, France\label{addr2}}
\maketitle

\begin{abstract}
We predict the sterile neutrino spectrum of some of the key solar nuclear reactions and discuss the possibility of 
these being observed by the next generation of solar neutrino experiments.  
By using an up-to-date standard solar model with good agreement with current helioseismology and solar neutrino flux data sets, 
we found that from solar neutrino fluxes arriving on Earth only
3\%-4\% correspond to the sterile neutrino.  The most intense solar sources of sterile neutrinos
are the $pp$ and $^7Be$  nuclear reactions with a total flux of $2.2\times 10^{9}\;{\rm cm^2 s^{-1}}$ and  
$1.8\times 10^{8}\;{\rm cm^2 s^{-1}}$, followed by the  $^{13}N$ and $^{15}O$ nuclear reactions  with a total flux of $1.9\times 10^{7}\;{\rm cm^2 s^{-1}}$ and  $1.7\times 10^{7}\;{\rm cm^2 s^{-1}}$.  Moreover, we  compute the sterile neutrino spectra of the nuclear proton-proton nuclear reactions -- $pp$, $hep$ and $^8B$  and  the carbon-nitrogen-oxygen -- $^{13}N$, $^{15}O$ and $^{17}F$  and the spectral lines of $^7Be$. 
\end{abstract}


\section{Introduction\label{sec-intro}}
The standard neutrino flavour oscillation model with only three neutrinos has been very successful in explaining most of observational properties of neutrino fluxes. Nevertheless, the significant improvement on the experimental methods of neutrino detection during the past fifteen years have revealed the existence of disagreements between the standard model of neutrino oscillations  and experimental data. 
These discrepancies usually refereed to as {\it neutrino anomalies} have been reported in a number of different type of experiments: the Liquid Scintillator Neutrino Detector~\citep[LSND;][]{2001PhRvD..64k2007A}, the Booster Neutrino Experiment,~\citep[MiniBooNE (first phase of BooNE);][]{2010PhRvL.105r1801A,2013PhRvL.110p1801A},
reactor experiments~\citep{2011PhRvD..83g3006M}  and 
the gallium experiments~\citep{2011PhRvC..83f5504G}.
Several generalizations of the standard neutrino flavour oscillation model have been proposed in the literature to explain these neutrino
anomalies~\citep[e.g.,][]{2008PhR...460....1G,2016EPJA...52...87M}.
Among other solutions the most successful one is  
 the  prediction of the existence of a new type of neutrinos~\citep{1962PThPh..28..870M}, known as
 {\it sterile neutrinos}. However, this experimental hint that can be explained by the existence of sterile neutrinos has been challenged by more recent measurements made by the 
 NEOS Experiment~\citep{2017PhRvL.118l1802K},  the  Super-Kamiokande detector~\citep{2015PhRvD..91e2019A}
 and the IceCube Collaboration~\citep{2017PhRvD..95k2002A}. Although the parameter space that allows the sterile neutrinos to explain the neutrino anomalies has been significantly reduced, their existence can still be accommodated within the current set of observations.

\smallskip
Sterile neutrinos are very well-motivated elementary particles and can be obtained from a trivial extension of the standard model of particle physics~\footnote{In the remainder of the article, the standard model will always refer to the standard model of particle physics if not stated otherwise.}. Their name reflects the fact that by definition sterile neutrinos are not affected by any of the known forces of nature except gravity, since these particles carry no charges and are singlets under all gauge groups of the standard model. This type of neutrinos, unlike other particle candidates, provides an unified description of three major problems in the framework of the standard model: the origin of the neutrino masses and the flavour oscillation model~\citep{2017JCAP...01..025A},  the absence of primordial anti-matter in the Universe and the existence of dark matter~\citep{2012PDU.....1..136B}.  Although many of the properties of this class of neutrinos is fixed by the generalized standard model, other properties of sterile neutrinos remain undefined, like their masses.
Actually, all sorts of theoretical arguments have been proposed to explain the existence of sterile neutrinos with a very large range of mass scales: from neutrinos with very light masses much smaller that  eV~\citep{2004PhRvD..69k3002D} up to very heavy neutrinos with masses of $10^{16}$ GeV~\citep{1977PhLB...67..421M}.
In recent years, particular attention has been dedicated to the sterile neutrino candidates with masses of a few eV, hundred keV  and a few GeV~\citep{1980PhLB...96..159S}  as several experiments are already obtaining data or are under construction with the goal of probing the existence of sterile neutrinos on these mass   ranges. 

\smallskip
The exact number of sterile neutrinos is not known, but even the simplest of the neutrino flavour oscillation models in which only one extra sterile neutrino $\nu_s$ is added up to the standard active  neutrinos $(\nu_e,\nu_{\tau},\nu_{\mu})$ is able to resolve many of the known experimental discrepancies -- the neutrino anomalies. This model is the so-called 3+1 neutrino flavour oscillation model in which the generalized mass matrix leads to the mixing of 4 neutrinos $(\nu_e,\nu_{\tau},\nu_{\mu},\nu_s)$.
This basic model  presents very compelling results compatible with
experimental data~\citep{2015PhRvD..91i5023B,2011PhRvD..83k3013P}.  Following
the literature~\citep[e.g.,][]{2015PhRvD..91i5023B}, in our study we will  focus on the 3+1 sterile neutrino in which two additional parameters  are included to characterize the sterile neutrino flavour oscillations: the mass spilling $|\Delta m^2_{41}|$  and the mixing angle $\theta_{14}$.

\smallskip
The mass spilling and the mixing angle (as defined in the 3+1 flavour neutrino oscillation model) that have been estimated by different research groups using several observational data sets and different fitting techniques, found very similar values for both quantities. A  global analysis of the neutrino oscillation data in the 3+1 parameter space made by~\citet{2013PhRvD..88g3008G} found that $\Delta m^2_{41}$ could take a value between  $0.82$--$ 2.19\; eV^2$ at 99.7\% confidence level. Similarly,~\citet{2017PhRvL.118l1802K}
making an independent analysis  have obtained identical values for $|\Delta m^2_{41}|$, specifically mass values in the range $0.2$ -- $2.3\;eV^2$. The determination of  $\theta_{14}$ by different research groups leads also to similar values. For instance,~\citet{2015PhRvD..91i5023B} and~\citet{2017PhRvL.118l1802K}, found for the upper limit of $\theta_{14}$ the values:
$10.6^o(0.1855)$ and 
$9.2^o(0.1582)$.  These values for $\theta_{14}$ differ only slightly from the first determination made by~\citet{2011PhRvD..83k3013P}
that found 
$11.7^o(0.204)$. Conveniently, for our calculations
we choose for the fiducial value of  $\theta_{14}$  the most recent 
determined value:  $9.2^o(0.1582)$ or $\sin^2{(2\theta_{14})}=0.1$~\citep{2017PhRvL.118l1802K}.

\begin{figure}
$$\qquad$$		
\centering 
\includegraphics[scale=0.45]{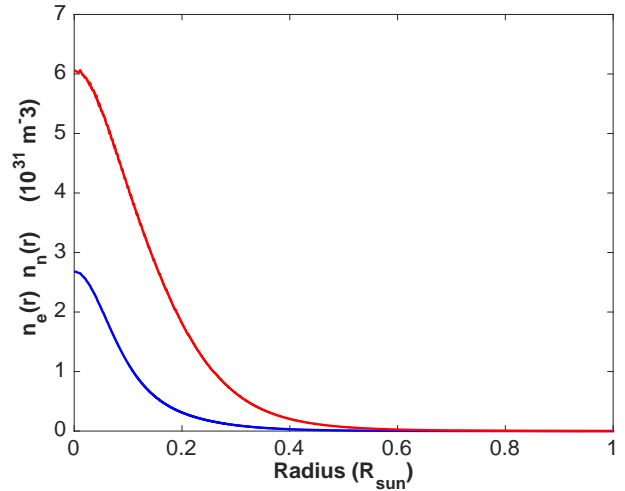}
\caption{The plasma of the Sun's core is made of
		ions and free electrons. The plot shows the variation of the number density of electrons $n_e(r)$ (red curve)
		and neutrons $n_n(r)$ (blue curve). In the computation of this quantities we used the solar standard model which is in
		excellent agreement with the helioseismic data.
		See~\citet{2013MNRAS.435.2109L} for details.}
	\label{fig:densities}
\end{figure}  

\smallskip
Since sterile neutrinos do not take part in the weak interaction,
the most obvious way to observe their effect is via their mixing with the active neutrinos.  Therefore, with the goal of  determining the sterile neutrino properties,  in this study we predict the spectra of sterile and electron neutrinos coming from the Sun, on the hope this will contribute to their discovery by "direct" detection by one of the ongoing experiments, the ultimate silver bullet that will prove their existence.  Specifically, we compute the sterile neutrino and electron neutrino spectra of some of the key nuclear reactions occurring in the Sun's core.   
These two types of neutrino spectra, as we will discuss in this work, results from the conversion of electron neutrinos into sterile neutrinos and from the survival of electron neutrinos, due to their oscillation between the 4 neutrino flavours, as these particles propagate through space (vacuum and matter).
 Specifically, we predict the shape of the different solar sterile and electron neutrino spectra of the key  neutrino nuclear reactions of the proton-proton (PP) chain and carbon-nitrogen-oxygen (CNO) cycle occurring in the interior of the present Sun.  
The calculation will be made for the current solar standard model which used the
most  up-to-date physics (equation of state, opacities, nuclear reactions rates and microscopic diffusion of heavy elements). The details about the physics of the solar model used in this study can be found  in~\citet{2013MNRAS.435.2109L,2013ApJ...765...14L}. This solar  model
predicts the sound speed profile and solar neutrino fluxes in agreement with observational data and  similar to other models found in the literature~\citep[i.e.,][]{2010ApJ...715.1539T,2011ApJ...743...24S}.

\begin{figure}
$$\qquad$$		
\centering 	
\includegraphics[scale=0.40]{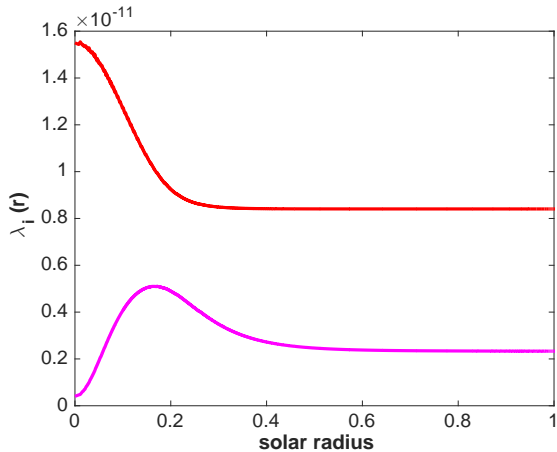}
\caption{
The level crossing scheme: The mass eigenvalues
$\lambda_1$ (red curve) and  $\lambda_2$ (magenta curve) 
as a function of the solar radius (cf. equation~\ref{eq:lambda}). These $\lambda$ values correspond to a neutrino with an energy $E=3.5 \; MeV$.}
\label{fig:levelcrossingscheme}
\end{figure}

\smallskip 
In the next section, we discuss in some detail the 3+1 sterile neutrino flavour oscillation model. In the following section, we present the spectra predictions for sterile and electron neutrinos of several of the key solar nuclear reactions. In the last section, we discuss the results and their implications for the next generation of sterile neutrino experiments. 
  
\begin{figure}[!t]
$$\qquad$$	
\centering 	
\includegraphics[scale=0.40]{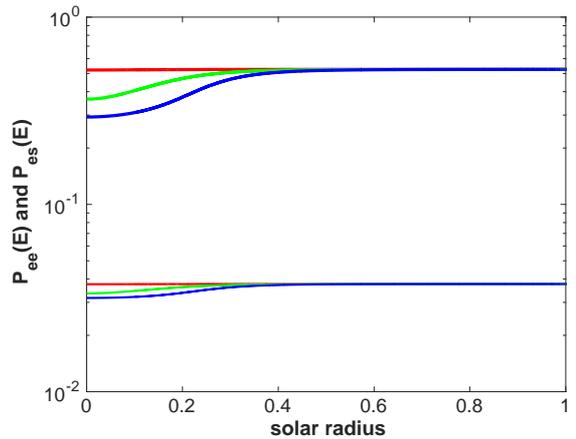}	
\caption{
The neutrino probability functions:  $P_{es}(E)$ 
corresponds to the probability of conversion of
$\nu_e$ to $\nu_s$ neutrinos given by equation~\ref{eq:Pessunrad}
(bottom set of curves); and  $P_{ee}(E)$  corresponds to  the survival probability of $\nu_e$  given by equation~\ref{eq:Peesunrad} (top set of curves). The red, green and blue curves correspond to neutrinos with energy of $0.1$,  $5$ and  $15$  MeV, which correspond to typical values of neutrino energies produced inside the Sun's core. In the computation of these quantities we used the standard solar model (see~\citet{2013MNRAS.435.2109L} for details).}
\label{fig:PePsrad}
\end{figure}     
    
\begin{figure}[!t]
$$\qquad$$	
\centering 	
\includegraphics[scale=0.40]{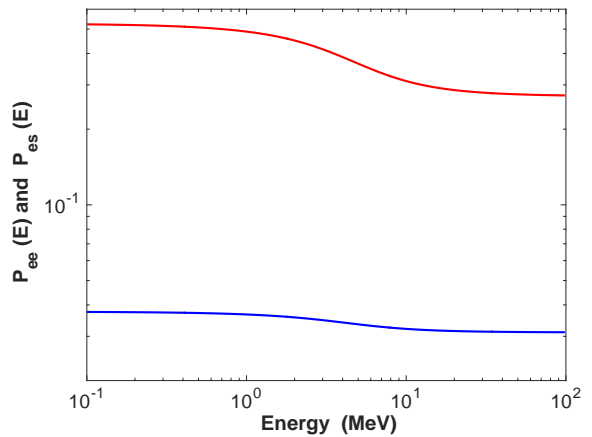}	
\caption{The neutrino probability functions as a function of 
the neutrino energy: $P_{es}(E)$  
corresponds to the probability of conversion of $\nu_e$  to $\nu_s$ neutrinos (blue curve); $P_{ee}(E)$ corresponds to the survival probability of $\nu_e$ neutrinos (red curve). See Figure~\ref{fig:PePsrad} for details. }\label{fig:PePsE}
\end{figure}       
     
\begin{figure}
$$\qquad$$	
\centering 
\includegraphics[scale=0.45]{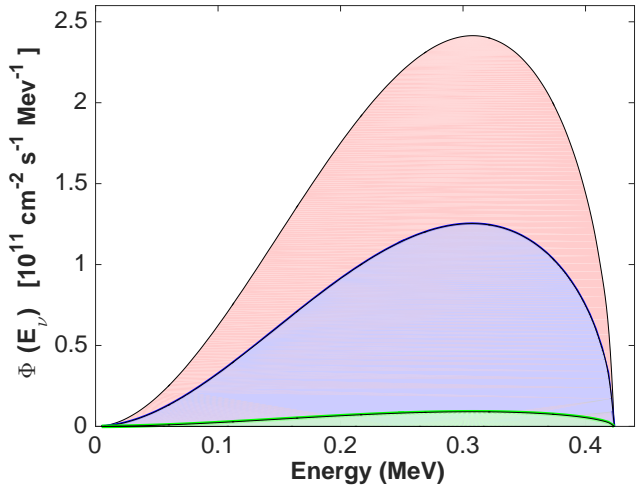}
\caption{Neutrino spectrum of $pp$ nuclear reaction: 
the $\Phi_{t} (E)$ (red area) is  the electron neutrino 
spectrum inside the Sun.  $\Phi_{ee}^{\odot} (E)$ (blue area) 
and $\Phi_{es}^{\odot} (E)$  (green area)  are the survival electron and sterile solar neutrino spectra for  the PP neutrinos  $3+1$ neutrino flavour oscillation model.}
\label{fig:nupp}
\end{figure}  

\section{Neutrino flavour oscillation models}
\label{sec-NPSC}

\begin{figure}[!t]
$$\qquad$$		
\centering 
\includegraphics[scale=0.45]{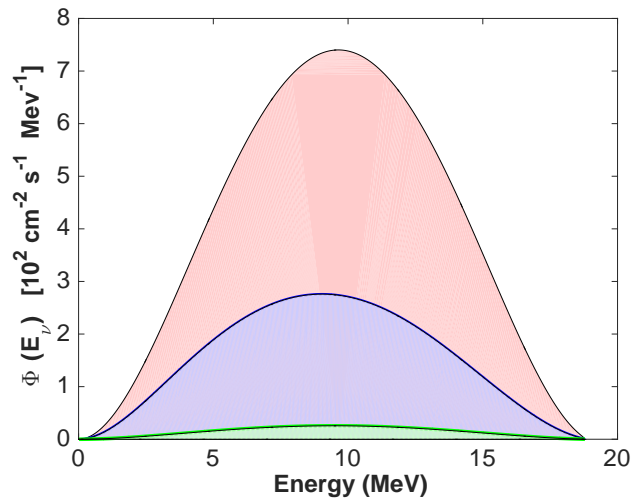}
\caption{Neutrino spectrum of $hep$ nuclear reaction.
The colour scheme is the same used in Figure~\ref{fig:nupp}.  }	
\label{fig:nuhep}
\end{figure}  

This study focuses on the simple and popular $3+1$  neutrino
flavour oscillation model with 4 neutrinos: $\nu_e$, $\nu_\mu$ ,$\nu_\tau$ and $\nu_s$. The addition of an extra (sterile) neutrino leads to a new flavour oscillation component on the $3+1$ neutrino model defined by a new mass splitting 
$|\Delta m^2_{41}|$ and a new mixing angle $\theta_{14}$.
As discussed previously,  a recent fit of the experimental data 
to the $3+1$ neutrino model 
done by~\citet{2015PhRvD..91i5023B}
found $\sin^2{\theta_{14}}=0.034$ (or $\theta_{14}\sim 10.6^o (0.1855)$), $\Delta m^2_{21}= (7.5\pm 0.14)\times 10^{-5}{\rm eV^2}$,  $\sin^2({\theta_{12}})=0.300\pm 0.016 $ and  $\sin^2{\theta_{13}}=0.03$.  In our calculations of the 3+1 neutrino oscillation model we will use these parameters. However the value of $\theta_{14}$  is replaced by a recent estimation of  $9.2^o(0.1582)$ obtained by~\citet{2017PhRvL.118l1802K}.
This choice of $\theta_{14}$ does not change our conclusions  since the difference between the two values is small to affect our results. 
It is important to highlight that the experimental data used by~\citet{2015PhRvD..91i5023B} 
for the previous fits comprises  data
from solar detectors like SNO, Super Super-Kamiokande, Borexino, Homestake, and Gallium experiments and  reactors data from the very long baseline KamLAND. Effectively, the inclusion of more experimental data, like the data sets coming from short baseline detectors such as the  DAYA-BAY,  RENO, and Double Chooz reactor experiments changes only slightly the value of
$\sin^2{\theta_{14}}$ and once again its magnitude variation is too small to significantly affect our results and 
conclusions~\citep{2015PhRvD..91i5023B}. Actually, although 
$\sin^2{\theta_{14}}$ has changed over the years, its present value 
is only slightly different from the value originally found  by~\citet{2011PhRvD..83k3013P}: $\sin^2{\theta_{14}}=0.041$.
Moreover, we observe that the $|\Delta m^2_{41}|$ determined  by~\citet{2017PhRvL.118l1802K} to be in the mass range of $\sim 0.2$ -- $2.3\;eV^2$ is much larger than the mass splittings, $|\Delta m^2_{21}|$ and $|\Delta m^2_{32}|$ estimated to be of the order of $10^{-5} eV^2$ and  $10^{-3} eV^2$~\citep{2017PhRvL.118l1802K,2016NuPhB.908..199G}.  
We note that in the 3+1 neutrino model discussed in our 
study only the mixing angle $\theta_{14}$ is considered to
have a value different from zero~\citep{2015PhRvD..91i5023B,2011PhRvD..83k3013P}.   
We stress this is a very good approximation, since a fit to the 3+1 neutrino model (with more mixing angles), that besides the new mixing angle $\theta_{14}$ also has  additional mixing angles like $\theta_{24}$ and $\theta_{34}$ (associated with other possible flavour  oscillations), found relatively small values for $\theta_{24}$ and $\theta_{34}$~\citep{2013JHEP...05..050K}. Therefore setting these last mixing angles to zero will not affect our results and conclusions. The experimental facts discussed above justify the analytical neutrino oscillation model that we will discuss in the next subsection. 
 
 \begin{figure}[!t]
 $$\qquad$$		
 \centering 
 \includegraphics[scale=0.45]{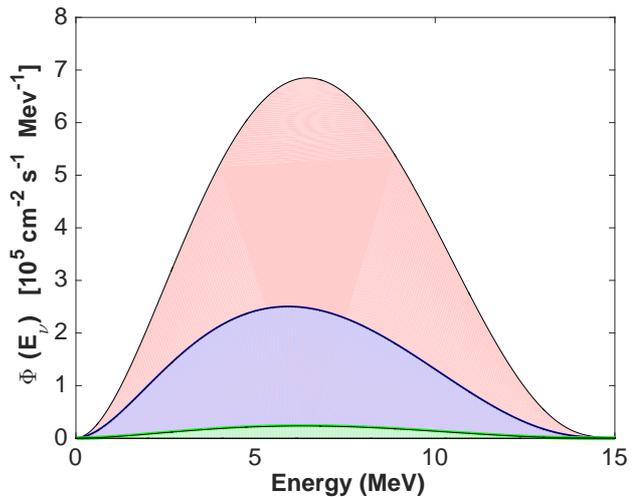}
 \caption{Neutrino spectrum of $^8B$ nuclear reaction.
The colour scheme is the same used in Figure~\ref{fig:nupp}.  }	
 \label{fig:nub8}
 \end{figure}  

\smallskip 
Finally, we point out that our $3+1$  neutrino model has parameters (excluding $\Delta m_{41}$ and $\theta_{41}$)
very similar to the ones found for the standard $3$  neutrino flavour oscillation model. For reference, a recent estimation of parameters made by~\citet{2016NuPhB.908..199G} using the  standard $3$ flavour neutrino oscillation model to fit the combined experimental data set obtained  from atmospheric experiments, solar detectors and particle accelerators found the following set of parameters:
$ \Delta m_{31}^2\sim 2.457 \pm 0.045\; 10^{-3} eV^2 $ 
(or $ \sim -2.449 \pm 0.048\; 10^{-3} eV^2 $) , 
$ \Delta m_{21}^2 \sim 7.500 \pm 0.019\; 10^{-5} eV^2$, 
$ \sin^2{\theta_{12}}=0.304 \pm 0.013 $  ,$ \sin^2{\theta_{13}}=0.0218 \pm 0.001 $,  $\sin^2{\theta_{23}}=0.562 \pm 0.032 $ and $\delta_{CP}=2\pi/25\; n$ with $n=1,\cdots,25$. 
This set of parameters  of values is also confirmed by other authors~\citep{2014JHEP...11..052G,2015JHEP...09..200B}.

\begin{figure}[!t]
$$\qquad$$		
\centering 
\includegraphics[scale=0.45]{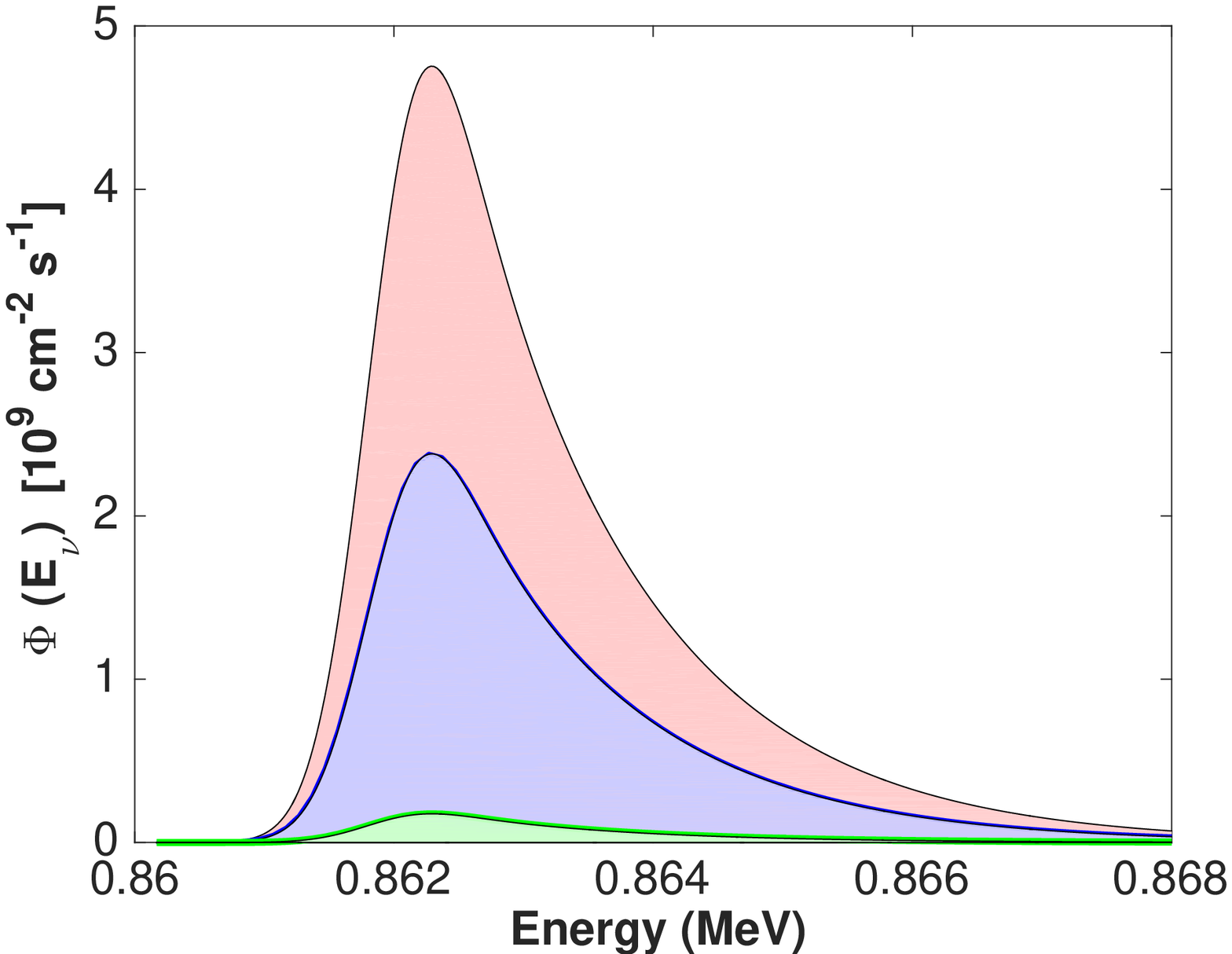}
\includegraphics[scale=0.45]{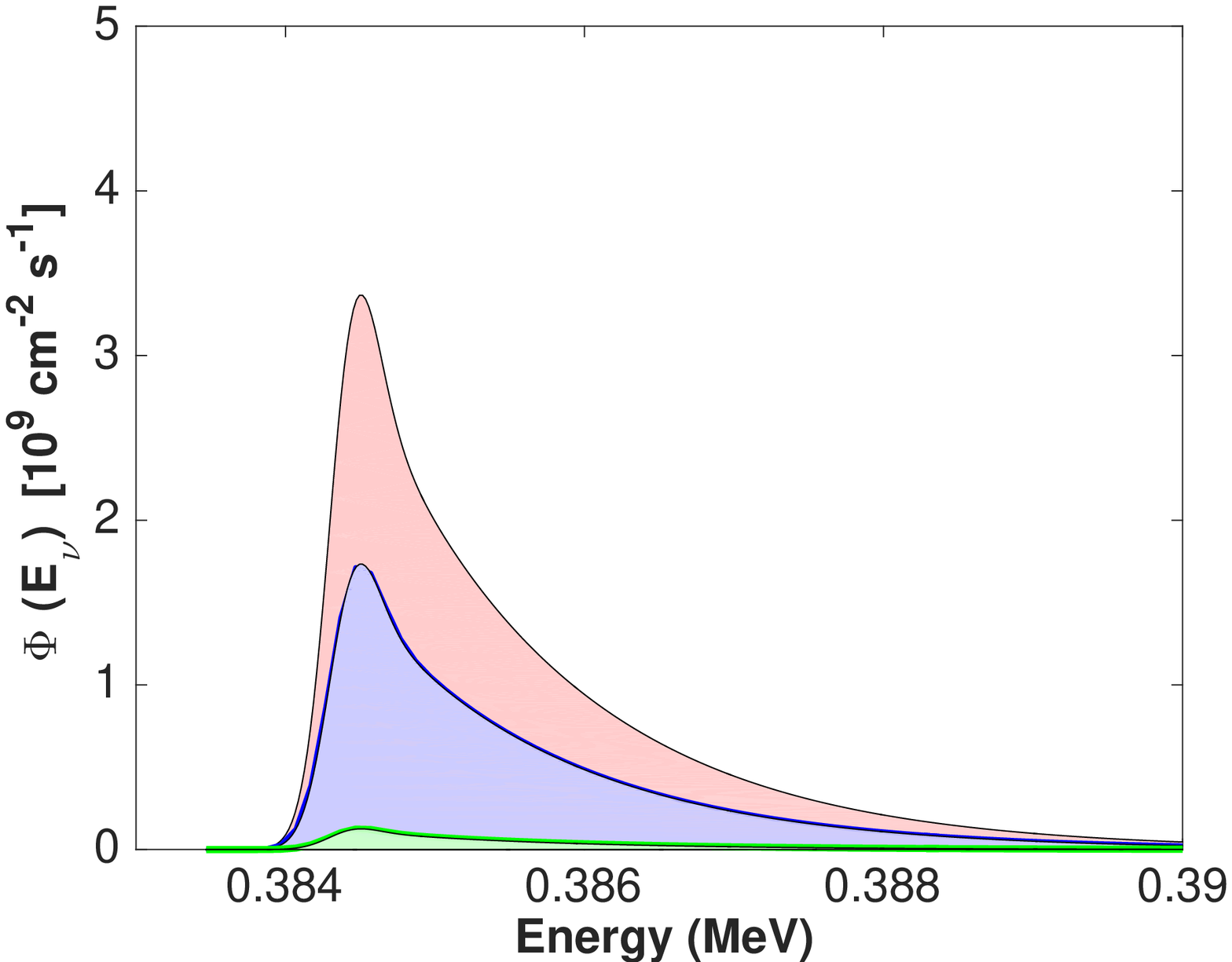}
\caption{
Neutrino of the two line spectra of $^7Be$ nuclear reaction.
The colour scheme is the same used in Figure~\ref{fig:nupp}.  }
\label{fig:nuBe7}
\end{figure}

\subsection{The standard and  3+1 neutrino flavour models}

The neutrino oscillation model with a fourth sterile neutrino $\nu_s$ 
is such that the neutrino flavours $(\nu_e, \nu_\mu, \nu_\tau, \nu_s)$ and	the mass eigenstates $(\nu_1, \nu_2, \nu_3, \nu_4)$ are connected through
a $4\times 4$ unitary mixing matrix $U$. We choose
for $U$ the parametrization proposed by~\citet{2011PhRvD..83k3013P} 
which is optimized to study the solar neutrino sector. It is worth highlighting that 
this peculiar parametrization is somehow between an admixture of
flavour eigenstates and matter eigenstates, as discussed  by~\citet{2005NuPhB.708..215C}.  Such $U$ formulation ensures this 3+1 neutrino model has the required sensitivity to the flavour-type and mass-type admixtures needed to study the solar data sector. In particular, this 3+1 mixing matrix is such that the standard neutrino model
corresponds to $\theta_{14}=0$.
A detailed account of the properties of this model can be found  
in~\citet{2011PhRvD..83k3013P}. In particular, the electron neutrino $\nu_e$ oscillation with the  $(\nu_1,\nu_2)$ mass eigenstates is only slightly affected by the decoupled  $(\nu_3,\nu_4)$ mass eigenstates for which the associated mixing angles are very small. In such  neutrino model, the MSW effect is computed in the hierarchical limit 
$\Delta m^2_{21}/2E\ll \Delta m^2_{31}/2E \ll \Delta m^2_{41}/2E$
where $E$ is the energy of the neutrino, an $\Delta m^2_{i1}=m^2_{i}-m^2_{1}$ where $m_i$ is the mass of each neutrino. 
This 3+1 neutrino model is an obvious generalisation  of the standard
(3 active) neutrino model~\citep{1989RvMP...61..937K}. In both cases neutrino oscillations are approximated by an  effective 2 neutrino oscillation model. As a consequence the $\nu_3$ and $\nu_4$ evolve independent of each other and
completely independent of the doublet: $(\nu_1,\nu_2)$.

\subsection{The eigenvalues of the mass matrix in 3+1 model}

In the Sun, as the experimental data on neutrino fluxes is limited, it is sufficient to represent the survival or conversion of neutrino flavor, 
using a two neutrino flavour oscillation model~\citep{1989RvMP...61..937K}.  The neutrinos emitted by the solar nuclear reactions are typically of low energy,
since these neutrinos have always an energy smaller than  20 MeV. 
Accordingly, some of the properties of solar neutrinos can be well understood by means of a two active neutrino flavour model: one electron neutrino $\nu_e$ and a second active neutrino, which we choose to be $\nu_\mu$, since the $\nu_\tau$ mixing angle is very small. Another way to interpret this approximation,  is to assume that $\nu_\mu$ now means the generic active $\nu_{\mu\tau}$,  if not stated otherwise.
In this much simpler 2 (active)+1 (sterile) neutrino flavour model ($\nu_e,\nu_{\mu},\nu_s$), the properties of the sterile neutrino $\nu_s$ can be computed as an approximation to  the ($\nu_e,\nu_{\mu},\nu_{\tau},\nu_s$) neutrino model.  
Accordingly, this 2+1 neutrino model has the following active mass eigenstates $\nu_1$ and $\nu_2$, as well as a sterile mass eigenstate\footnote{The option to use the subscript $4$ for the mass eigenstate of the sterile neutrino is motivated by the fact of maintaining a simple correspondence between the 2+1 and 3+1 neutrino flavour models.} $\nu_4$~\citep{2004PhRvD..69k3002D,2005NuPhB.708..215C}. 
The 2+1 flavour model in the absence of sterile neutrinos have the  following
eigenvalue  expressions~\citep{2004PhRvD..69k3002D}. If $\lambda_i$ ($i=1,2$) are the correspondent eigenvalues of the mass matrix, it reads
\begin{eqnarray}
\lambda_i=\frac{M^2}{4E}+\frac{V_e+V_{n}}{2}\mp	\sqrt{\left(\Delta_{c}-\frac{V_e-V_{n}}{2}\right)^2
+\Delta_{s}^2},
\label{eq:lambda}
\end{eqnarray}
 where $M^2=m_1^2+m_2^2$,
$\Delta{c} \equiv \Delta m_{21}^2\cos{(2\theta_{21})}/(4E)$
and $\Delta{s} \equiv \Delta m_{21}^2\sin{(2\theta_{12})/(4E)}$,
$V_e=\sqrt{2}G_F(n_e-n_n/2)$ and $V_{n}=-\sqrt{2}G_Fn_n/2$,
where    $G_F$ is the Fermi constant and $n_e(r)$ and $n_n(r)$ are the profile of electron and neutron
densities of the solar interior.
The electron density $n_e(r)$ is given by $N_o \rho(r)/\mu_e(r)$, where $\mu_e(r)$ is the mean molecular weight per electron, $\rho(r)$ is the density of matter in the solar interior, and $N_o$ is Avogadro's number. 
The density of neutrons $n_n(r)$  is computed in a similar manner like $n_e(r)$ where  the mean molecular weight per electron
$\mu_e(r)$  is replaced by the mean molecular weight per neutron
$\mu_n(r)$.

\smallskip
Figure~\ref{fig:densities} shows the current abundances
of electrons and neutrons  inside the Sun (as a function of the solar radius) for its present age as predicted by an up-to-date standard solar model~\citep{2013MNRAS.435.2109L}. The large amount of $n_n(r)$ in the center of the Sun results from the fact that the Sun's core is made of more than 70\% of $^4He$. This occurs as a result of this element being continuously produced (nucleosynthesis) by the  conversion of 4 protons (0 neutrons) into an ion of $^4He$ (2 neutrons). This process has being occurring since our star arrived to the main sequence 4.5 Gyr ago and will continue to do it for the next 5 Gyr (until the Sun leaves the main sequence). As a consequence, at the present age the core of the Sun has a larger amount of neutrons than the more external layers, leading to a decrease of $n_n(r)$ from the center towards the surface of the star. This important point is illustrated in Figure~\ref{fig:densities} where we show for electrons and neutrons the relative density and mean molecular weight per particle inside the Sun. 

\smallskip
For illustrative purposes we assume that  $m_i^2 \approx \Delta m^2_{i1} $ (with $i\ne 1$). Specifically, $m_1^2 \approx 0 $,
$m_2^2 \approx 7.5 \times 10^{-5}\; eV^2 $
(with $\theta_{12}\approx 0.6$),   $m_3^2 \approx 2.4 \times 10^{-3}\; eV^2 $ (with $\theta_{13}\approx 0.13$), 
and  $m_4^2 \approx 1.0 \; eV^2 $
(with $\theta_{14}\approx 0.16$). The $\lambda_i$ ($i=1,2$) can be seen in the figure~\ref{fig:levelcrossingscheme}. In this model for which $m_1<m_2 <m_3 <m_4 $, there is only a relevant resonance in the system associated with 1-2 level crossing. The 1-2 resonance energy is given by the expression
$E_{12}= \Delta m_{21}^2{\cos{(2\theta_{12})}}/({2\sqrt{2}G_Fn_e})$.
As the electron neutrino moves outward, the electronic density decreases
and the neutrino eventually go through the resonance region $E_{12}(r)$ 
and beyond. However, since the electron density $n_e(r)$ in the Sun's core varies slowly, i.e., $dn_e(r)/dr$ is small, the propagation is adiabatic, the
neutrino state will remain in the same mass eigenstate,  then the $\nu_e$ will emerge associated with $\nu_1$.  The flavor composition of $\nu_2$ is now determined by the mixing angle in the vacuum and the $\nu_2$ is now mostly $\nu_{\mu\tau}$. 

We notice that for other neutrino models the eigenvalues  can have additional resonances. For instance,~\citet{2004PhRvD..69k3002D} 
have shown that for a 2+1 neutrino flavour model 
($\nu_e,\nu_{\mu\tau},\nu_s$) with  $m_1 < m_4 < m_2$,  the sterile neutrino level $\lambda_4$ crosses $\lambda_1$. In this model, the $\lambda_2$ essentially decouples, therefore it is not affected by the s-mixing. However, in such a neutrino model,  due to the  variation of the $n_e(r)$ in the Sun's core, 
the eigenvalues have two resonances: one related with the active neutrinos,  the 1-2, i.e., the same resonance mentioned previously; and a new resonance, the  $1-4$,  that corresponds to $\lambda_4$ crosses $\lambda_1$  in a region at the right of the 1-2 resonance\footnote{As explained by~\citet{2004PhRvD..69k3002D},
in  principle  $\lambda_4$ could cross $\lambda_1$ twice: one  above and another below the 1-2 resonance.}. However, this case does not occur in our study, since in our 2+1 model $\lambda_4$ is much larger than  $\lambda_1$ and $\lambda_2$ and consequently does not cross any of them.

\subsection{The conversion of sterile neutrino in electronic neutrinos }
 
The goal of this work is to predict the spectra of sterile and electronic neutrinos emitted by the Sun for which their specific shapes strongly depend on the properties of the plasma of the solar interior.  As such, in the following
we start to compute the probabilities of  conversion of electron neutrino to  sterile neutrino  $P_{es}$ $[\equiv P (\nu_e \rightarrow \nu_s]$ arriving on  Earth. Equally we also compute the survival probability of electron
neutrino $P_{ee}$ $[\equiv P (\nu_e \rightarrow \nu_e]$.
In this 3+1 neutrino flavour oscillation model like in the standard 
3 flavour oscillations~\citep[e.g.,][]{2010LNP...817.....B}, the neutrino 
oscillates between the available flavours due to oscillations
in vacuum and matter~\citep[Mikheyev-Smirnov-Wolfenstein (MSW), e.g.,][]{1978PhRvD..17.2369W}. The survival and conversion probabilities of a four-flavour neutrino oscillation can be reduced to a modified three-flavour and two-flavour  neutrino oscillation model. These simplifications results from the difference 
of magnitude between mass splitings and mixing angles.
A detailed discussion about the neutrino flavour oscillation model can be found in~\citet{2008PhR...460....1G}.  

Since the evolution of neutrinos in matter is  adiabatic, for that reason their contribution for  $P_{es}$ and $P_{ee}$ can be cast in simple expressions~\citep{2011PhRvD..83k3013P}. Following, the usual parametrization of the neutrino mixing matrix this leads to the following expression for $P_{es}$ (probability conversion from  $\nu_e$ to $\nu_s$):
\begin{eqnarray}
P_{es}=s^2_{14}c^2_{14}c^4_{13}\bar{P}^{2\nu}_{ee}+s^2_{14}c^2_{14}s^4_{13}
+s^2_{14}c^2_{14},
\label{eq:Pessunrad}
\end{eqnarray}
where $c_{ij}=\cos{\theta_{ij}}$ and $s_{ij}=\sin{\theta_{ij}}$. 
Equally the survival probability of electron neutrinos $P_{ee}$  
reads  
\begin{eqnarray}
P_{ee}=c^4_{14}c^4_{13}\bar{P}^{2\nu}_{ee}+c^4_{14}s^4_{13}+s^4_{14}.
\label{eq:Peesunrad}
\end{eqnarray} 
In the previous equations $\bar{P}^{2\nu}_{ee}$ is the electron neutrino survival probability in a two-flavour neutrino oscillation model. It reads
\begin{eqnarray}
\bar{P}^{2\nu}_{ee}=c^2_{12}(c^m_{12})^2+s^2_{12}(s^m_{12})^2,
\end{eqnarray} 
where $\theta^{m}_{12}$ is the  matter  angle that depends equally of the fundamental parameters of neutrino flavour oscillations and the properties of solar plasma.  $\theta^{m}_{12}$  can be equally determined from one of the following expressions:  
\begin{eqnarray}
\cos{(2\theta^m_{12})}
=\frac{\cos{(2\theta_{12})}-V_m(r)}{
	\sqrt{\left[\cos{(2\theta_{12})}-V_m(r)\right]^2+\left[\sin{(2\theta_{12})} \right]^2}},
\end{eqnarray} 
or 
\begin{eqnarray}
\sin{(2\theta^m_{12})}=
\frac{\sin{(2\theta_{12})}}{
	\sqrt{\left[\cos{(2\theta_{12})}-V_m(r) \right]^2+\left[\sin{(2\theta_{12})} \right]^2}}.
\end{eqnarray} 
In the previous expressions $V_m(r)$ is the effective matter potential given by
\begin{eqnarray}
V_m(r)=c^2_{13}\left(c^2_{14}+f_m(r)s^2_{14}\right)
\;\frac{2E\sqrt{2}G_Fn_e(r)}{\Delta m^2_{21}}.
\label{eq:Vm}
\end{eqnarray} 
The factor $f_m(r)$ in a function that gives the ratio of the neutron density  $n_n(r)$ in relation to the electron density  $n_e(r)$ at each radius of the solar interior, reads
\begin{eqnarray}
f_m(r)=\frac{1}{2}\frac{n_n(r)}{n_e(r)}.
\end{eqnarray} 

The variation of $V_m(r)$ (cf. equation~\ref{eq:Vm}) depends of  
the radial variation of neutrons as well as electrons inside the star.
We notice, that in the particular case that  $\theta_{14}=0$,  the $3+1$ neutrino oscillation model becomes identical to the standard 3 neutrino oscillation model (for oscillations in vacuum and matter). 
Consequently, in this case $ V_m(r)$ is reduced to a much simple expression~\citep{2010LNP...817.....B}: $V_m(r)=c^2_{13}\;{2E\sqrt{2}G_Fn_e(r)}/{\Delta m^2_{21}}$.  
These quantities (cf. equation~\ref{eq:Vm})  have a major impact on the formation of the sterile and electronic neutrino spectra. 
Figures~\ref{fig:PePsrad} and~\ref{fig:PePsE} show the $P_{ee}$ and
$P_{es}$  for the current standard solar model.

\begin{figure}[!t]
$$\qquad$$		
\centering 
\includegraphics[scale=0.45]{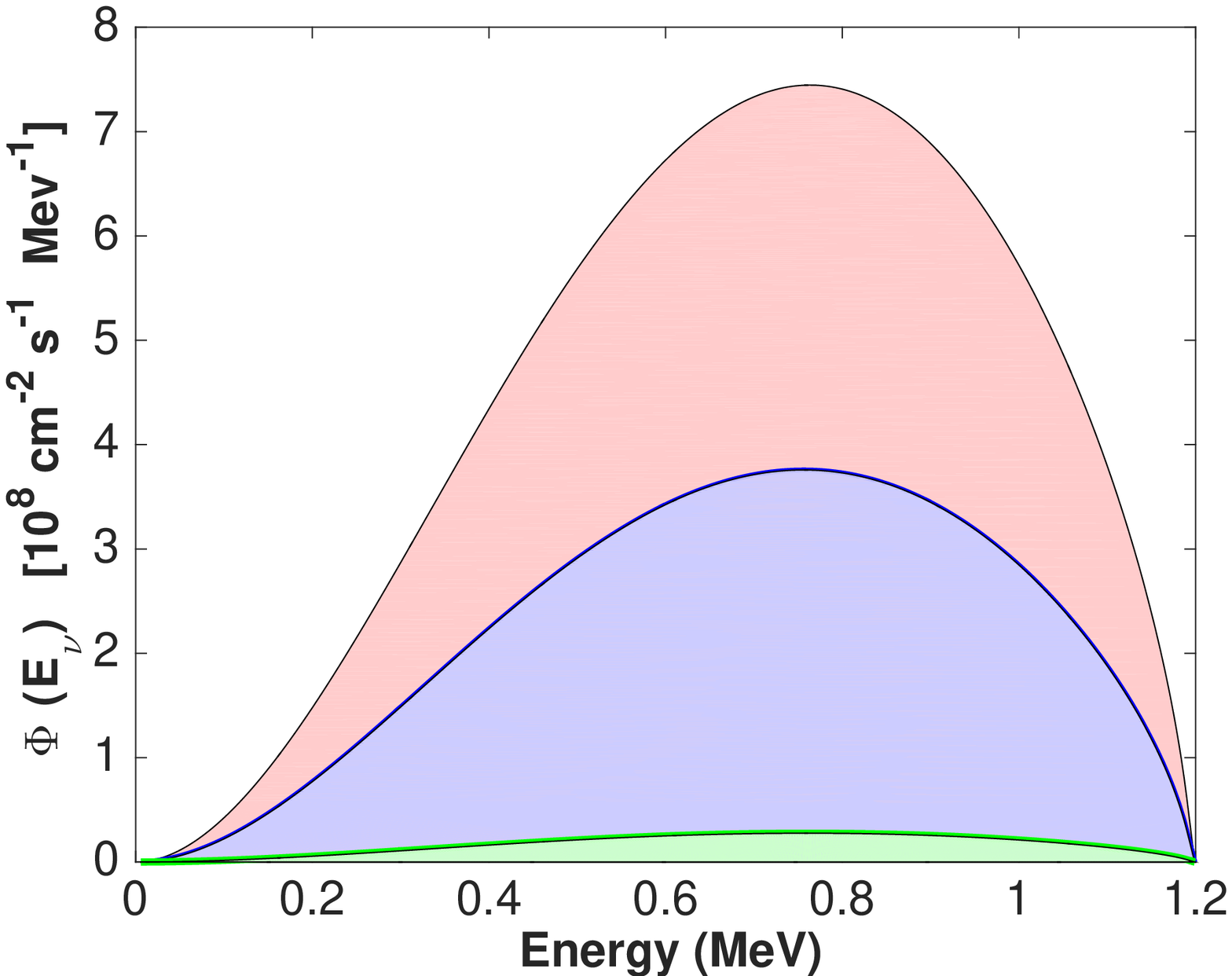}
\includegraphics[scale=0.45]{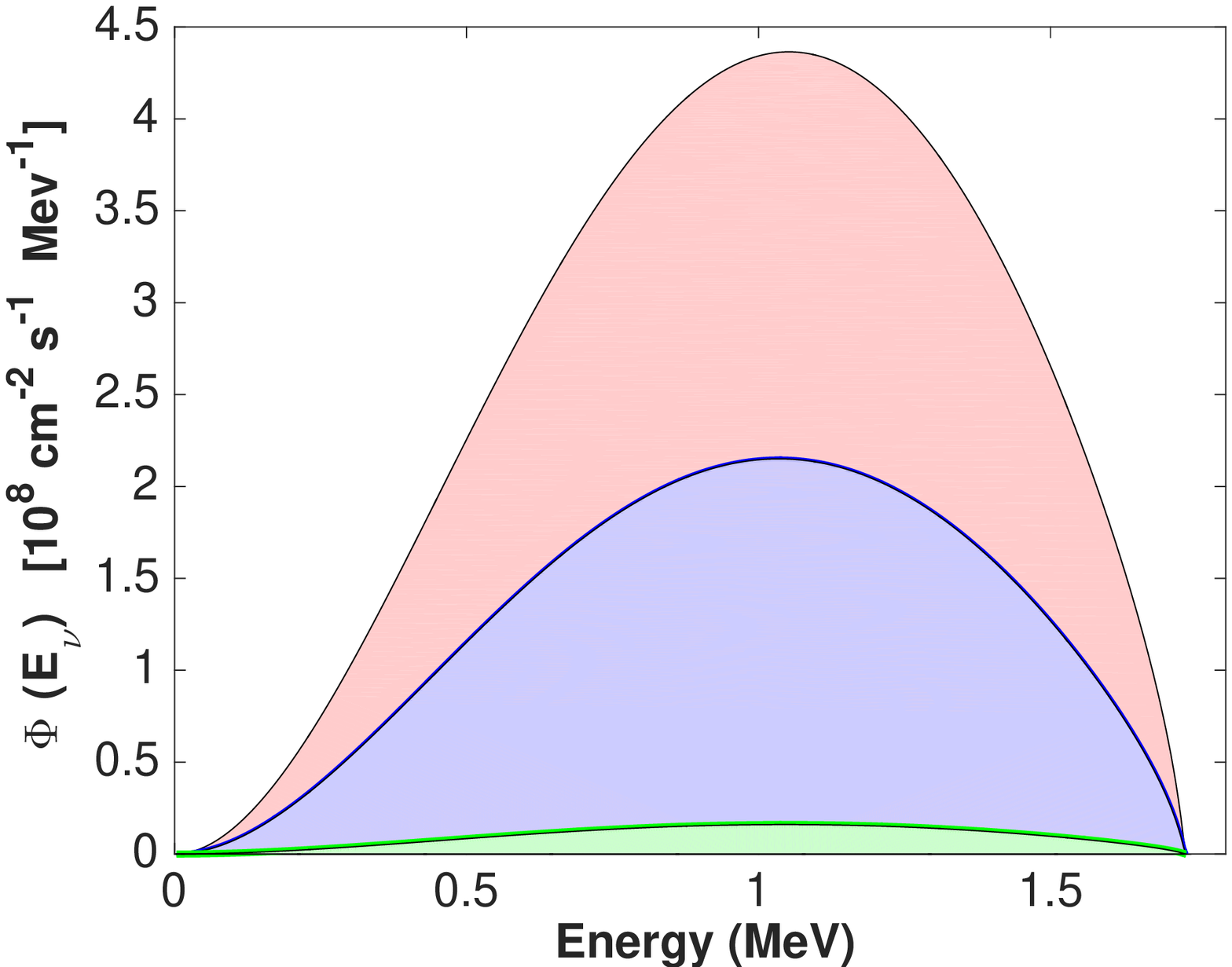}
\includegraphics[scale=0.45]{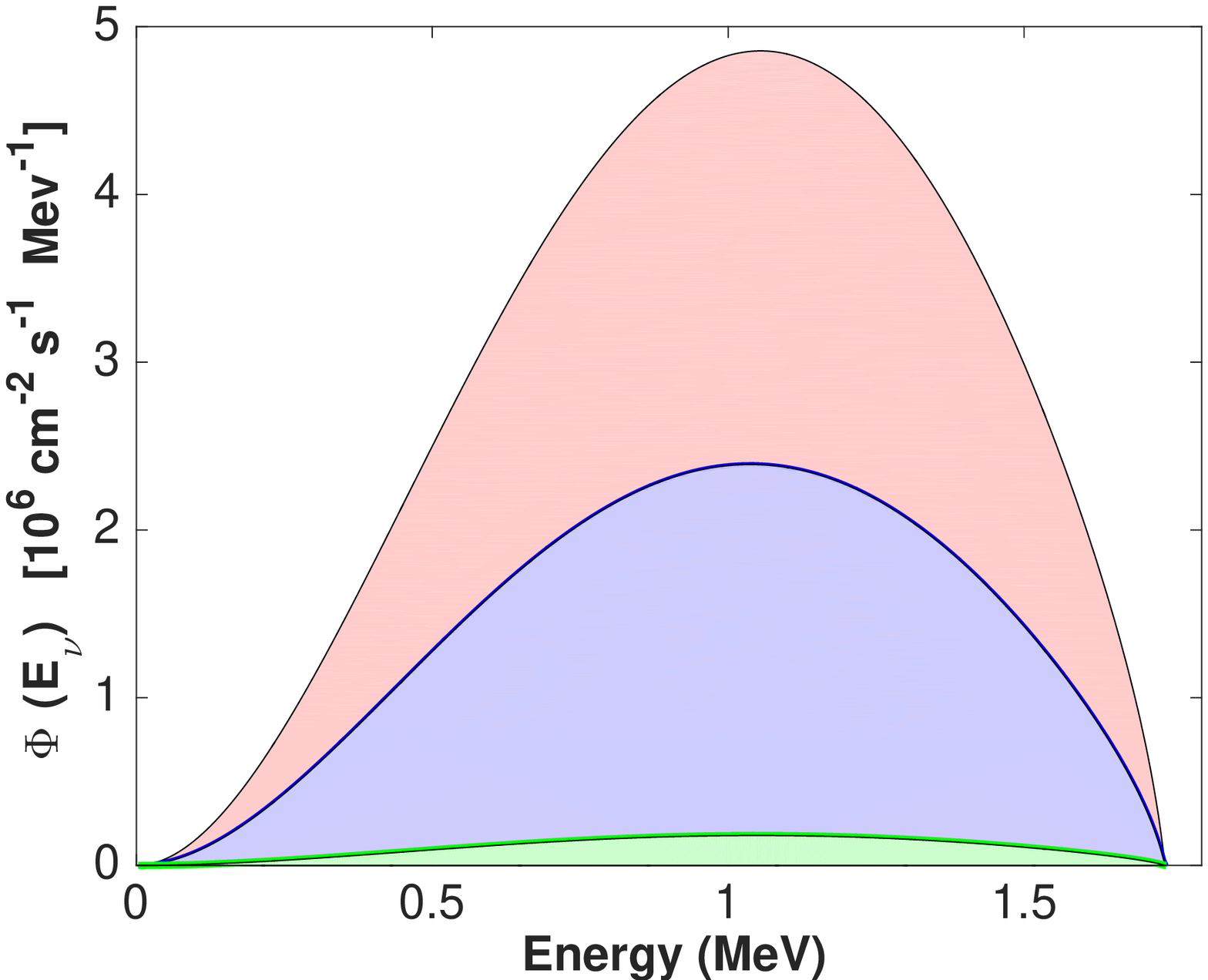}
\caption{The solar neutrino spectra for CNO neutrinos:
Neutrino spectra of the $cno$ nuclear reactions.	
$N13\nu$ (top), $O15\nu$ (middle) and	$F17\nu$ (bottom). 	 
The colour scheme is the same used in Figure~\ref{fig:nupp}.  }
\label{fig:nucno}
\end{figure}  
 
\subsection{Emission of electron neutrino spectra from solar nuclear reactions}

Since neutrinos are produced by nuclear reactions (either the PP chains and the CNO cycle) in different locations of the Sun's core, consequently, the survival probability of electron  neutrinos will depend on the distance of the nuclear reaction to the center of the Sun~\citep{2013PhRvD..88d5006L,2017PhRvD..95a5023L}.
This will also affect indirectly the probability of conversion of electron neutrinos into sterile neutrinos. The average probability  $P_{ee,k}$  of electron neutrinos of energy $E$ produced by a $k$ nuclear reaction region is given by
\begin{eqnarray} 
P_{ee,k} (E) = 
A_k^{-1} \int_0^{R_\odot} P_{ee} (E,r)\phi_k (r) 4\pi \rho(r) r^2 dr, 
\label{eq:Pnuej}
\end{eqnarray}  
where  $ A_k $ is a normalization constant thait is equal to $\int_0^{R_\odot}\phi_k (r) 4 \pi \rho (r) r^ 2  \;dr $. 
$ \phi_k (r) $ 
is the electron neutrino emission function for the
following $k$ nuclear reactions: $pp$, $hep$, $^8B$, $^7Be$, $^{13}N$, $^{15}O$ and $^{17}F$. 
In this study, we consider that all neutrinos produced in the solar nuclear reactions are of  electron flavour as predicted by standard nuclear physics, therefore, the local density of neutrons only affects the $P_{ee,k} (E)$ by affecting the flavour of electron neutrinos due to the existence of a  new sterile neutrino $\nu_s$. It is important to note that $ \phi_k (r) $, the neutrino source of each nuclear reaction, is sensitive to the local values (radius $r$) of the temperature,  molecular weight, density, neutron and electronic densities.

\smallskip
In a 3+1 neutrino flavour oscillation model, as neutrinos propagate in the Sun's interior, the neutrino flavour oscillate between the four flavour states  $\nu_e$, $\nu_\mu$, $\nu_\tau$  and $\nu_s$ either due to vacuum or matter oscillations. Although in outer space (between the Sun and the Earth) the flavour oscillations due to matter effects are expected to be small, the high density of the solar interior medium will enhance the scattering of neutrinos with other  elementary particles, like electrons and neutrons leading to the increase of the probability of the neutrino oscillating between  different flavour states. The particular distribution of electrons and neutrons in the solar interior leads to characteristic shapes of neutrino spectra (of electron neutrinos and sterile neutrinos) for each of the neutrino nuclear reactions occurring in the Sun's interior, as we will discuss in the next section
(see figures: \ref{fig:nupp} -- ~\ref{fig:nucno}).

\section{Prediction of the solar sterile and  electron neutrino spectra}
\label{sec-SEMNS}

\begin{figure}
$$\qquad$$	
\centering 
\includegraphics[scale=0.45]{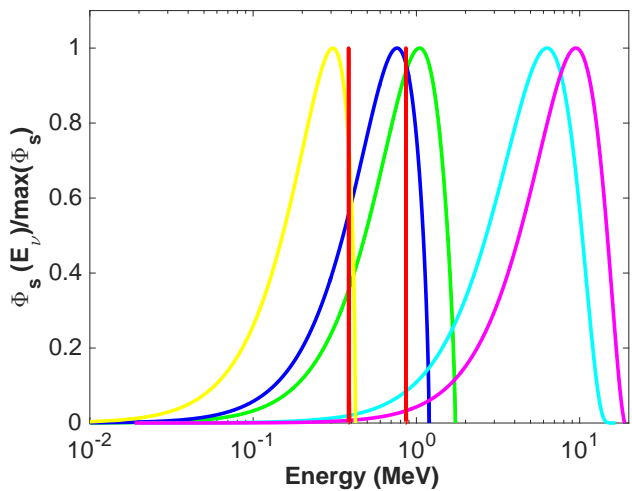}
\caption{Normalized sterile neutrino spectra for the Sun:
		$pp$ (yellow curve), $^8B$ (cyan curve), $Be7$ (red curve)
		$hep$ (magenta curve)	
		$N13$ (blue curve), $O15$  and	$F17$ (green curve).}
	\label{fig:nusterile}
\end{figure}  

\smallskip 
The spectrum of electron neutrinos from any specific nuclear reaction, like the $\beta$--decay processes occurring in the Sun's core are known to be essentially independent  of the properties of the solar plasma. Actually, the neutrino spectrum emitted by these nuclear reactions occurring inside the Sun are identical to the same spectra emitted by the seminar nuclear reactions tested on many laboratories on Earth. Therefore, it is reasonable to assume that the neutrino emission spectra is the same in both cases~\citep{2017PhRvD..95a5023L}. Since the $\beta$--decay processes only produce electron neutrinos, using the theory of neutrino flavour oscillation is possible to predict in detail the  spectrum shapes of sterile and electronic neutrinos arriving on Earth.

\begin{table} 
	\centering
	\caption{Solar neutrino sources of sterile neutrinos}
	\resizebox{8cm}{!}{  
	\begin{tabular}{lllll} 
		\hline
		\hline
		Neut.  &  $\Phi_t$
	   &    
	   $\Phi_e$ &    $\Phi_s$  \\
		source &  ${\rm  cm^{-2} s^{-1}} $ &  ${\rm  cm^{-2} s^{-1}} $ & ${\rm  cm^{-2} s^{-1}} $   \\
		\hline
		{\bf pp chain}  &   &  &   \\
		$pp$  &  $5.94\; 10^{10}$ &$3.09\; 10^{10}\;(52\%)$ &  $2.22\; 10^{9}\;(3.7\%)$  \\
		$hep$  &  $7.88\;10^3 $ &$2.88\; 10^{3}\;(37\%)$ &  $2.64\;10^2\;(3.4\%)$  \\
		$^8B$  &  $5.27\;10^6 $ &$1.91\; 10^{6}\;(36\%)$ &  $1.76\; 10^{5}\;(3.3\%)$ \\
		$^7Be\; \footnotesize{861.3\; keV}$   &  $4.75\;10^9$ &$2.38\; 10^{9}\;(50\%)$ &  $1.76\; 10^{8}\;(3.7\%) $ \\
		$^7Be\; \footnotesize{384.3\; keV}$   &  $4.75\;10^9 $ &$2.44\; 10^{9}\;(51\%) $ &  $1.77\; 10^{8}\;(3.7\%) $ \\
		\hline
		{\bf cno cycle}  &   &  &   \\
		$^{13}N$  &  $5.32\; 10^{8}$ &$2.69\; 10^{8}\;(51\%)$ &  $1.97\; 10^{7}\;(3.7\%)$ \\
		$^{15}O$  &  $4.49\; 10^{8}$ &$2.21\; 10^{8}\;(49\%) $ &  $1.65\; 10^{7}\;(3.7\%)$ \\
		$^{17}F$  &  $5.01\; 10^{6}$ &$2.47\; 10^{6}\;(49\%)$ &  $1.84\; 10^{5}\;(3.7\%)$ \\
		\hline
		\hline
	\end{tabular}}
\noindent
\flushleft{\scriptsize \hspace{0.2cm} The electronic neutrino fluxes   ($\Phi_t$) predicted for the current solar} 
 \vspace{-0.3cm}
\flushleft{\scriptsize \hspace{0.2cm} model as in reference~\citet{2013MNRAS.435.2109L}.}
	\label{tab:nufluxes}
\end{table}   

\smallskip 
Here we make predictions of  the sterile neutrino spectra of the main nuclear reactions occurring in the Sun, in the hope that this can be used as a guide for the future generation of neutrino detectors, which could help to establish (or rule out) the existence of sterile neutrinos. Once that the neutrinos emitted in the solar nuclear reactions are all of electron flavour $\nu_e $, than the existence of other neutrino flavours in the solar neutrino spectrum (as it is possible to observe on solar neutrino detectors) is uniquely related with the nature of neutrino flavour oscillations, which leads to the convection of $\nu_e$ on other neutrino flavours: $\nu_\mu $,  $\nu_\tau $ and $\nu_s$.
Accordingly, the electron neutrino spectrum of the $k$ nuclear reaction is defined as  $\Phi_{k,t}(E)$  (with the subscript $t$ referring to the emission  electron neutrino spectrum of the nuclear reaction $k$),   
and $\Phi_{k,s\odot}(E)$  and  $\Phi_{k,e\odot}(E)$  are the
sterile neutrino spectrum and the electron neutrino spectrum
arriving on Earth.  The function $\Phi_{k,t}(E)$ corresponds to the electron neutrino spectrum of a specific nuclear reaction computed theoretically, and also in many cases measured in the laboratory. The $^8$B neutrino spectrum is a very good example.  The $^8$B neutrino spectrum~\citep{1986PhRvC..33.2121B} has been  shown to fit the experimental data with a high degree of accuracy~\citep[e.g.,][]{2006PhRvC..73b5503W}. 
Hence,  the sterile neutrino spectrum  $\Phi_{k,s\odot}(E)$  is computed as 
\begin{eqnarray} 
\Phi_{k,s\odot}(E)=P_{es,k}(E) \Phi_{k,t}(E),
\label{eq-Phies}
\end{eqnarray}
where $k$ is equal to  one of the following values:
$pp$, $hep$, $^8B$, $^7Be$, $^{13}N$, $^{15}O$ and $^{17}F$. 
Similarly the electron neutrino spectrum  $\Phi_{k,e\odot}(E)$  is given by  
\begin{eqnarray} 
\Phi_{k,e\odot}(E)=P_{ee,k}(E) \Phi_{k,t}(E).
\label{eq-Phiee}
\end{eqnarray}
Figures~\ref{fig:nupp},~\ref{fig:nuhep},~\ref{fig:nub8},~\ref{fig:nuBe7} and ~\ref{fig:nucno} show the conversion of the emission neutrino
spectra $\Phi_{k,t\odot}(E)$ in the two other spectrum flavours   $\Phi_{k,e\odot}(E)$ and $\Phi_{k,s\odot}(E)$.
Moreover, table~\ref{tab:nufluxes} shows the total neutrino fluxes computed
for the present-day standard solar model~\citep{2013MNRAS.435.2109L}.   
For each nuclear reaction, $\Phi_t$ is the total neutrino flux emitted by  the nuclear reaction, and  $\Phi_e$ and $\Phi_s$  correspond to the flux fractions of $\Phi_t$ that arrive at the solar detector as electron neutrinos  or sterile neutrinos. 

\begin{figure}
$$\qquad$$	
\centering 
\includegraphics[scale=0.45]{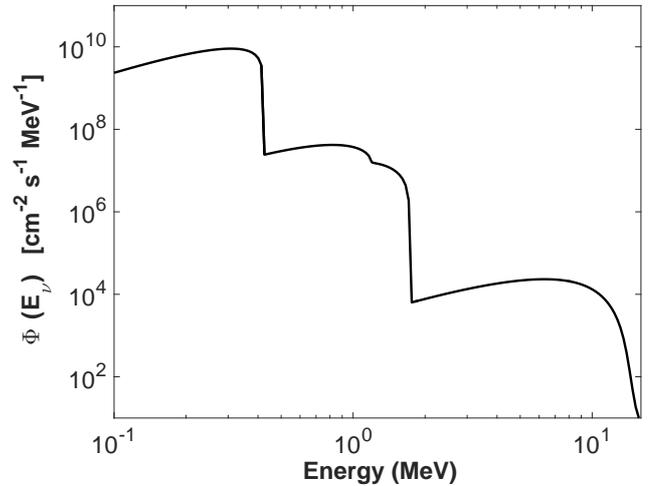}
\caption{The total sterile neutrino energy spectrum predicted by our standard solar model. The individual solar neutrino fluxes are shown 
in Table~\ref{tab:nufluxes}. }
\label{fig:nusteriletot}
\end{figure}

\smallskip
We start to notice that due to neutrino oscillations the electron neutrino spectra arriving on Earth detectors for all the solar nuclear reactions (considered in this study) are very different from the neutrino spectra emitted by the solar nuclear reactions (without neutrino oscillations).  In particular, it is possible that 
the next generation of solar neutrino detectors will be able to measure $\Phi_{k,e\odot}(E)$. 
The sterile neutrino spectrum of all nuclear reactions have very small amplitudes and have identical spectral shapes although occurring at different neutrino energy ranges.  
In table~\ref{tab:nufluxes} we show the predicted neutrino fluxes for
the leading solar nuclear reactions for the current solar standard model as well as the predicted ratios of sterile neutrino fluxes and 
electronic neutrino fluxes. The small values of sterile neutrinos are due to the very small mixing angle $\theta_{14}$. Only $4\%$ of neutrinos arriving on Earth have a sterile flavour. This value is almost constant across all solar nuclear reactions. Nevertheless there is a small increase for the sterile neutrinos from the $pp$ reaction and a small decrease for sterile neutrinos from the $^8B$ reaction. This is a consequence of the fact that the $pp$ and $^8B$ nuclear reactions
produce neutrinos with the smallest and the largest energies 
which due to the specific dependence sterile neutrino flavour
oscillations with the local distribution of electrons and neutrons
in the solar core leads to this specific variations. It is interesting to note that all sterile neutrinos in the solar neutrino spectrum have
unique shapes related with the specific solar nuclear reactions on its
origin. For instance, the two spectral lines of $^7Be$ have strong asymmetry in their shape which could be helpful to look for hints related with sterile neutrinos in the solar neutrino spectrum. The strongest source of solar  sterile neutrinos are the $pp$ and $^7Be$  nuclear reactions with a total flux of $2.22\times 10^{9}\;{\rm cm^2 s^{-1}}$ and  
$1.76\times 10^{8}\;{\rm cm^2 s^{-1}}$, followed by the  $^{13}N$ and $^{15}O$ nuclear reactions  with a total flux of $1.97\times 10^{7}\;{\rm cm^2 s^{-1}}$ and  $1.65\times 10^{7}\;{\rm cm^2 s^{-1}}$ (cf. Table~\ref{tab:nufluxes}).  
Figure~\ref{fig:nusteriletot} shows the cumulative sterile neutrino spectrum 
coming from the Sun. The maximum emission of sterile solar neutrinos corresponds to an energy of 0.307 MeV and a neutrino flux $9.0\times 10^{9}\;{\rm cm^2 s^{-1}}$. 

\section{Conclusion} 
\label{sec-DC}

Sterile neutrinos are among the best candidates to resolve some of the
leading problems in particle physics. At the same time these are very
viable candidates to dark matter.
Nevertheless, the hint of its existence has been questioned by a recent group of new experiments. 
The final proof beyond doubt of sterile neutrino existence will pass by its direct detection.

\smallskip
 
In this study we have computed for the first time the sterile spectrum of several key solar nuclear reactions. We found that
from the solar neutrino flux arriving on Earth only $3\%-4\%$ 
corresponds to sterile neutrinos. As expected,  the strongest source
is the  $pp$ nuclear reaction with a total flux of $2.22\times 10^{9}\;{\rm cm^2 s^{-1}}$ followed by the  $^7Be$  nuclear reaction with a total flux one order of magnitude smaller.  Equally, we have also predicted in detail the spectral shape of the sterile neutrino spectrum associated to the different solar nuclear reactions. In particular as shown in figure~\ref{fig:nusterile} the shape of these sterile neutrino spectra are quite distinct.   

Final point: the Sun produces a significant amount of  sterile neutrino on the spectral range from 10 KeV to 10 MeV,
mostly coming for the  $pp$ nuclear reaction (in the neutrino energy range
from 0.01 MeV up to 0.4 MeV), and $^7Be$  (spectral lines  0.861 MeV and 0.384 MeV)  that can affect experiments which are looking for relic (non-relativist) active and sterile neutrinos like  {\sc Katrin } and {\sc Ptolemy} Collaboration~\citep{2017EPJC...77..410B,2013arXiv1307.4738B}.   
Indeed several experiments based in different principles have been proposed to detect such relic neutrinos (both components active and sterile neutrinos)~\citep{2008JPhCS.110h2014C,2010PhRvD..81k3006A}: electron-capture of decaying nuclei~\citep{2011JCAP...08..006L,2011PhLB..695..205L}, annihilation of high-energy cosmic neutrinos at the Z-resonance,  atomic de-excitation~\citep{2005PhRvD..71h3002B,2015PhRvD..91f3516Y} and neutrino capture using radioactive beta-decaying nuclei~\citep{1962PhRv..128.1457W}. The later seems to be  the most promising technique with a few experiments already planned to that end like the {\sc Katrin } and {\sc Ptolemy}
projects~\citep{2017EPJC...77..410B,2017NIMPA.848..127D,2016PhRvD..94k6009H,2013arXiv1307.4738B}. In an typical $\beta$--decay process like the one expected
to be found in the  {\sc Ptolemy} experiment, the relic neutrinos (cosmic neutrino background) will be shown as two independent kinks in the continuous $\beta$-spectrum,  one due to the active neutrinos and a second small peak at higher energy associated to the  admixture of sterile neutrinos~\citep{2014JCAP...08..038L}. The exact location of the energy peaks on the spectrum will depend on the masses of the active and sterile neutrinos.  
  
\smallskip
In conclusion, we have computed the total flux of sterile neutrinos
emitted by the Sun, as well as the solar sterile neutrino spectrum,
this could be helpful in looking for sterile signatures by the next generation of  solar neutrino detectors.

\medskip

\noindent
{\bf Acknowledgements}
The author thanks the Funda\c c\~ao para a Ci\^encia e Tecnologia (FCT), Portugal, for the financial support to the Center for Astrophysics and Gravitation (CENTRA), Instituto Superior T\'ecnico,  Universidade de Lisboa, 
through the Grant No. UID/FIS/ 00099/201.  Moreover, the author also acknowledges the support given by the Institut d'Astrophysique de Paris during his sabbatical visit.
%


\end{document}